\documentclass[twocolumn,showpacs,preprintnumbers,amsmath,amssymb]{revtex4-2}

\DeclareUnicodeCharacter{03B2}{\ensuremath{\upbeta}}
\usepackage{upgreek} 

\usepackage{siunitx} 
\usepackage{physics} 
\usepackage{mleftright} 
\usepackage{csquotes} 
\usepackage{graphicx} 

\newcommand{\cf}{\ensuremath{C\! F} }%

\begin{document}
\title{Exciting the Higgs mode in a strongly-interacting Fermi gas by interaction modulation}

\author{Andreas Kell, Moritz Breyer, Daniel Eberz, Michael K{\"o}hl}
	\affiliation{Physikalisches Institut, University of Bonn, Wegelerstra{\ss}e 8, 53115 Bonn, Germany}

\begin{abstract}
We study the Higgs mode of a strongly-interacting Fermi gas in the crossover regime between a fermionic and bosonic superfluid. By periodically modulating the interaction strength of the gas, we parametrically excite the Higgs mode  and study its resonance frequency and line width as a function of both interaction strength and temperature. We find that the resonance frequency at low temperature agrees with a local-density approximation of the pairing gap. Both frequency and line width do not exhibit a pronounced variation with temperature, which is theoretically unexpected, however, in qualitative agreement with a different recent study.
\end{abstract}

\maketitle


A strongly-interacting Fermi gas in  the crossover regime between a Bardeen-Cooper-Schrieffer (BCS) superconducting regime and a Bose-Einstein condensate (BEC) of dimers is a prime example of a tuneable many-body quantum system. The ability to experimentally vary the interaction strength of the gas allows for exploring the crossover from a primarily fermionic (BCS) to a primarily bosonic (BEC) system. It is well established that all across this BEC-BCS crossover there is a phase transition between a  normal and a superconducting state \cite{Leggett1980,Nozieres1985,regal_observation_2004,Zwierlein2005,Randeria2014,Link2023,Eberz2023}. However, it remains important to understand the fundamental excitations of the superfluid.

\begin{figure}
	\includegraphics[width=.8\columnwidth]{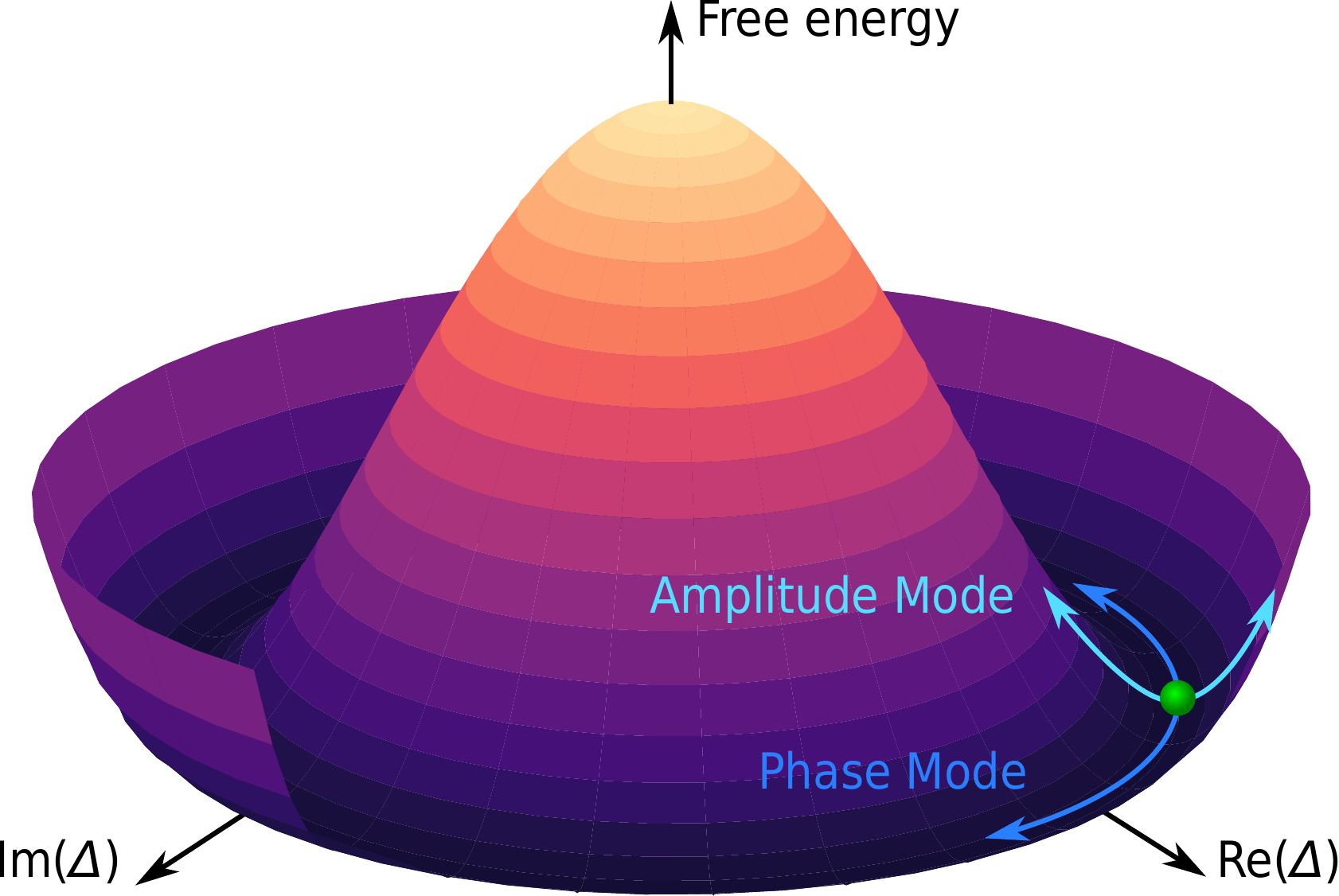}
	\includegraphics[width=\columnwidth]{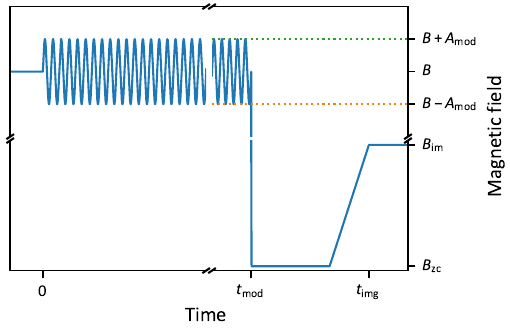}
	\caption{
		{\bf a)} Free energy diagram of the order parameter $\Delta$  together with the two fundamental excitations: (1) the phase/Goldstone mode  along circumference of free energy minimum and (2) the amplitude/Higgs mode of the amplitude of the order parameter in the radial direction. {\bf b)} Experimental scheme for the excitation of the Higgs mode by interaction modulation. A modulation of an external magnetic field $B$ with amplitude $A_\text{mod}$ modulates the $s$--wave scattering length $a$ and excites the ensemble. After modulation for a duration of  $t_\text{mod}$, a rapid ramp towards the the BEC side converts Cooper pairs to low-momentum molecules, which are detected by absorption imaging.
	}\label{f:fig1}
\end{figure}

The important parameter for determining this phase transition is the superconducting gap $\Delta$. The superconducting gap  both sets the energy scale of the critical temperature (in weak coupling) and serves as the order parameter for the phase transition. It is a complex quantity with an amplitude and a phase, and crossing the  phase transitions brakes the symmetry of the order parameter related to the complex phase, see Figure\,\ref{f:fig1}a. The free-energy landscape of Figure\,\ref{f:fig1}a supports different possible excitations of the order parameter. They include the Bogoliubov-Anderson mode \cite{Bogoljubov1958,Anderson1958a,Anderson1958b, hoinka_goldstone_2017,biss_excitation_2022}, which corresponds to the Nambu-Goldstone phase mode at low energies, and the Higgs mode as the collective amplitude mode of the order parameter \cite{volkov_collisionless_1973}. However, while a Goldstone is always present, the stability of the Higgs amplitude mode requires additional symmetries to prevent a fast decay into phase modes \cite{pekker_amplitude_2014}. 
In the BCS limit, particle-hole symmetry decouples amplitude and phase modes such that the Higgs mode becomes observable \cite{Balseiro1980,Littlewood1981} and, recently, it has been found that the Higgs mode could be stable for strongly-correlated condensates \cite{Lorenzana2024}. The Higgs amplitude mode manifests itself as an oscillation of the amplitude of the order parameter with frequency $\omega_\text{H}=2 \Delta / \hbar$ \cite{volkov_collisionless_1973}.

The Higgs amplitude mode has been experimentally investigated in various  superconducting systems \cite{sooryakumar_raman_1980, matsunaga_higgs_2013,chu_phase-resolved_2020,Cea2016,Udina2022} and in ultracold atomic gases in the BEC-BCS crossover regime \cite{behrle_higgs_2018, Dyke2024}. In the latter case, a mode had been observed even at and around unitarity. Broadly speaking, there are two excitation schemes, which have been proposed and/or used to successfully excite the Higgs mode: parametric excitation by a periodic modulation close to the resonance frequency \cite{scott_rapid_2012,matsunaga_light-induced_2014,behrle_higgs_2018,Collado2018,Collado2021,lyu_exciting_2023,Barresi2023,Collado2023} or probing the response of the system following an impulsive excitation \cite{volkov_collisionless_1973,matsunaga_higgs_2013,Hannibal2015,hannibal_persistent_2018,Dyke2024}. Recently, the Higgs amplitude mode excited by an interaction quench was observed in a  unitarity gas by probing a small, homogeneous region inside the sample using Bragg spectroscopy \cite{Dyke2024}. These measurements were performed for different temperatures and, while higher temperatures clearly led to a decreased amplitude of the oscillations of the order parameter, the frequency of the  Higgs mode was unchanged. The latter finding is surprising since the  superconducting gap exhibits a pronounced temperature dependence.



Here, we report on a measurement of the Higgs mode, its linewidth, and its temperature dependence in the BEC-BCS crossover using parametric excitation. The parametric excitation could address different parameters, such as the density \cite{behrle_higgs_2018,lyu_exciting_2023} or the interaction strength \cite{scott_rapid_2012,bayha_observing_2020}, and we opt here for modulating the interaction strength. The key advantage of this method is that no third, final state is involved in the measurement unlike in the radiofrequency modulation technique \cite{behrle_higgs_2018} (involving a different spin state) and in Bragg spectroscopy technique \cite{Dyke2024} (involving different momentum states). This rules out possible contributions of final-state interactions or scattering onto the data. Moreover, as compared to the radiofrequency modulation technique, we have an improved spectroscopic resolution since all momentum states couple equally to a chosen drive frequency. A similar excitation method has previously been employed to study trap dynamics, the molecular bound state on the BEC side and pair breaking excitations \cite{greiner_probing_2005}.

Experimentally, we start from a balanced mixture of degenerate $^6$Li in the hyperfine-states $\ket{1}$ and $\ket{3}$ prepared in a gravity-compensated optical dipole trap with trap frequencies of $(\omega_x,\omega_y,\omega_z) = 2 \pi \times \left(69, 95, 138\right)\,\si{Hz}$ \cite{Link2023}.
The atom number $N_\sigma$ is between \num{6.9e4} and \num{2.4e5} per spin state, depending on the target magnetic field for which the samples are prepared.
We determine the density distribution of our samples from {\it in-situ} high-intensity absorption imaging and applying an inverse Abel transformation for reconstructuion of the density \cite{ku_revealing_2012}.
The density $n_\sigma$ at the trap center defines the Fermi energy $E_\text{F} = \left(6 \pi^2 n_\sigma\right)^{2/3} \hbar^2/\left(2 m\right)$ and is in the range of $E_\mathrm{F}/h=13...32$\,kHz.
From the density distribution far away from the center we determine the temperature by fitting it to the virial expansion of the equation of state, which yields temperatures in the range of $T/T_\mathrm{F}=\SIrange{0.08}{0.13}{}$, depending on the interaction strength of the gas. We can increase $T/T_\textrm{F}$ further by slowly compressing and quickly decompressing the trap for temperature dependent measurements \cite{Link2023}.

The interaction strength between the atoms is set by a Feshbach resonance \cite{zurn_precise_2013} at a magnetic offset field between 664\,G and 698\,G, which corresponds to a range of interaction parameters $1/\mleft( k_\text{F} a \mright)$ of \num{-0.19} to 0.49. Here, $k_\textrm{F}$ denotes the Fermi wave vector and $a$ is the $s$--wave scattering length. For our experiments, we have chosen the states $\ket{1}$ and $\ket{3}$ because they feature the least broad Feshbach resonance among the relevant Li hyperfine states so that it is technically easier to achieve a high amplitude for the interaction modulation.

The excitation and measurement scheme is visualized in Figure 1b. The interaction modulation is performed by a periodic variation of the magnetic field in the vicinity of the Feshbach resonance. During the modulation, the Higgs amplitude mode is excited and, over time, leads to a reduction of the order parameter \cite{scott_rapid_2012}.  Technically, the magnetic field modulation is performed by  an additional magnetic field with an amplitude $A_\textrm{mod}$ of up to \SI{2}{G} and a sinusoidal time dependence with a frequency range of roughly $\nu_\text{mod} = 1 \dots 100\, \si{kHz}$ \cite{kell_compact_2021}. Then, without delay, the condensate fraction is measured by employing the rapid ramp technique \cite{regal_observation_2004}.
As a measure for the order parameter we determine the condensate fraction. Therefore, our experimental implementation is very close to \enquote{protocol B} from ref.~\cite{scott_rapid_2012}, in which the authors argue that there is no requirement that the final ramp speed must be faster than the oscillations of the order parameter. This is different to Bragg spectroscopy \cite{Dyke2024}, where the pulse length must be considerably shorter than the Higgs oscillation period thereby limiting the spectral resolution of the measurement.
A consequence of our excitation and detection method is that the whole inhomogeneous density distribution of the trapped sample contributes to the measurement.

\begin{figure}
	\includegraphics[width=\columnwidth]{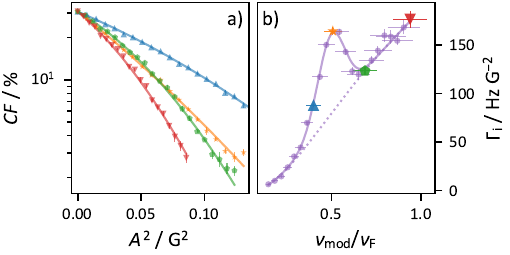}
	\caption{
		{\bf a)} Decay of the condensate fraction of a unitary Fermi gas as a function of modulation amplitude $A$ for modulation frequencies [a] 6.8\,kHz (blue triangle), [b] 8.6\,kHz (orange star), [c] 11.7\,kHz (green pentagon) and [d] 16\,kHz (red triangle). Solid lines are fits according to equation (1).
		\textbf{b)} Extracted values of the initial decay rate $\Gamma_\text{i}$ for different modulation frequencies $\nu_\text{mod}$. The specific data points extracted from a) have matching colors and symbols. The line is a fit to the data comprising of a 4\textsuperscript{th} degree polynomial background (dotted line) plus a Gaussian peak (solid line). $\nu_\mathrm{F}=E_\mathrm{F}/h$ denotes the Fermi frequency.
		Horizontal bars indicate systematic errors on $\nu_\mathrm{F}$, whereas vertical bars indicate statistical errors and are smaller than the symbols for most data points.
	}\label{f:cf_decay}
\end{figure}

For small depletion of the condensate, we observe that the condensate fraction \cf decays exponentially  with increasing modulation time $t_\text{mod}$ and increasing modulation amplitude squared $A_\text{mod}^2$ as described by the empirical relation $\cf = \cf_0 \exp\mleft( -\Gamma_\text{i} A_\text{mod}^2 t_\text{mod} \mright)$. Here, we have introduced the initial condensate fraction decay rate $\Gamma_\text{i}$, which quantifies the early response of the system at a given modulation frequency and takes the role of our main observable. The dependence on the amplitude squared is consistent with the behavior in ref.~\cite{greiner_probing_2005} and theoretical considerations on the BEC side \cite{plata_magnetic-modulation_2006}. We choose to keep the modulation duration constant at $t_\text{mod} = \SI{100}{ms}$ and vary the amplitude $A_\text{mod}$ to determine $\Gamma_\text{i}$. In order to account also for stronger depletion of the condensate fraction, we extend our fit formula by a correction factor $\gamma$ with
\begin{equation}
	\cf = \cf_0 \exp\mleft( -\Gamma_\text{i} A_\text{mod}^2 t_\text{mod} - \gamma [A_\text{mod}^2 t_\text{mod}]^2 \mright),\label{e:curvature}
\end{equation}
which reproduces the data very well, see Figure~\ref{f:cf_decay}a. We extract $\Gamma_\text{i}$ by fitting equation (1)  to the data for different modulation frequencies and show the spectrum of $\Gamma_\text{i}$ in Figure 2b. The spectrum features a peak on top of an increasing background, which we describe by a Gaussian plus a 4\textsuperscript{th} degree polynomial. We identify the symmetric peak with the collective excitation and attribute the background, among others, to incoherent pair-breaking excitations \cite{behrle_higgs_2018,Kurkjian2019}.


%
%

The spectrum is measured for the range of interactions  $1/\mleft(k_\text{F} a\mright)=-0.19...0.49$, and we show spectra in Figure 3. In Figure 3a, we show the dependence of the resonance on interaction strength and one can clearly observe how the peak in the modulation spectra shifts to higher frequencies as the coupling parameter $1/\mleft(k_\text{F} a\mright)$ of the Fermi gas is tuned from the BCS to the BEC side. In Figure 3b, we show the temperature dependence of the modulation spectrum at unitarity. For this analysis, we have set $\gamma=0$, which has allowed for a more robust fitting at smaller condensate fractions, where the signal-to-noise becomes lower. The possibility to conduct fits at samples with small condensate fractions has generally limited the highest measured temperature at unitarity.

\begin{figure}
	\includegraphics[width=\columnwidth]{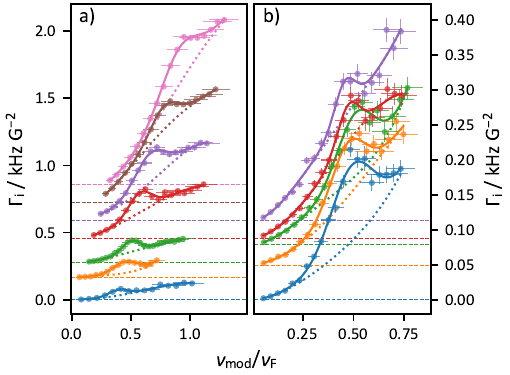}
	\caption{
		{\bf a)} Measured spectra for interaction parameters in the range from \num{-0.19} to 0.49 (bottom to top) for the lowest achievable temperatures. For better visibility, we introduce an offset (dashed) of $\left(1.5 / \mleft(k_\text{F} a\mright) + 0.28\right)\,\si{kHz G^{-2}}$, which is added to the data.
		The fits are a 4\textsuperscript{th} degree polynomial background (dotted) plus a Gaussian peak (solid).
		{\bf b)} Spectra at unitarity for different  temperatures in the range of $T/T_\textrm{F}= 0.08...0.13$ (bottom to top) and an offset (dashed) of $2 \left(T/T_\text{F} - 0.08\right)\,\si{kHz G^{-2}}$.
		Horizontal bars indicate systematic errors on $\nu_\mathrm{F}$, whereas vertical bars indicate statistical errors.
	}\label{f:spectrum}
\end{figure}

%
%

\begin{figure}
	\includegraphics[width=\columnwidth]{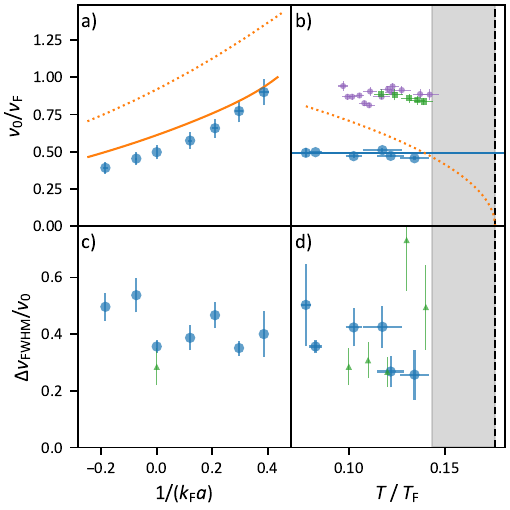}
	\caption{
		{\bf a)} Resonance position as function of interaction parameter. The solid line is the local-density weighted trap average of $2 \Delta/E_\mathrm{F}$ and the dotted line is the corresponding value at the center of the trap. 
		{\bf b)} Temperature dependence of the resonance position of the unitary gas. Our data (blue) show no significant variation and the solid blue line is the average of all data points. Green and purple data points are from reference \cite{Dyke2024} taken at the trap center.
		{\bf c)} Resonance width as a function of interaction strength (blue data points).  The data points are taken in a temperature range $T/T_\mathrm{F}=0.08...0.13$. Green data points are from reference \cite{Dyke2024} taken at the trap center.
		{\bf d)} Resonance width at unitarity as a function of temperature.
		The error bars in a) show the systematic uncertainty of the Fermi energy. Here, the statistical errors are smaller than the symbols. For b), c) and d) only statistical errors are displayed. 
	}\label{f:interaction_dependence}
\end{figure}

In the next step, we analyse both the center positions and the linewidths of the spectral features and compare to the theory of the Higgs mode. In Figure 4a, we show the  measured resonance frequency normalized to the Fermi frequency $\nu_\mathrm{F}=E_\mathrm{F}/h$ for different coupling strengths. In order to compare to theory, we perform a local-density approximation \cite{hannibal_persistent_2018,Musolino2021,Dyke2024} of the gap parameter \cite{haussmann_thermodynamics_2007} by weighting the local normalized Higgs frequency $\omega_\text{H}\mleft(r\mright)$ by $\Delta^2(r)$ assuming a Thomas-Fermi density profile, and we display this quantity as the solid line. The trap-averaged value gives an excellent agreement  with experimentally observed frequencies. For reference, we also display the value of $2\Delta/E_\mathrm{F}$ evaluated at the center of the trap as a dotted line.

In Figure 4b, we display the temperature dependence of the resonance frequency at unitarity. Notably, we do not observe any temperature variation. Note that we can measure only up to $T/T_\mathrm{F} \sim 0.15$ but not in the (grey-shaded) region up to the critical temperature because the condensate fraction is too small and we did not detect a resonance. The absence of a temperature dependence is in agreement with recent measurements of the Higgs mode using Bragg spectroscopy \cite{Dyke2024}, which we display in the graph as well. The difference between our measurements and those of ref. \cite{Dyke2024} is that our data are trap averaged whereas theirs are measured at the trap center. We scale their Fermi energy by the factor $\left(n_0/\bar{n}\right)^{2/3} \approx 0.97$ to account for a different definition of the Fermi energy. The weak temperature dependence could be related to the existence of a pseudo-gap state above the critical superfluid temperature, however, these findings warrant further investigation in the future, both theoretically and experimentally.

In Figures 4c and 4d, we show the full width at half maximum (FWHM)  of the spectral features. Our data could indicate a weak dependence on both interaction strength and temperature. 
Moreover, we also would expect a shortening of the lifetime with increasing temperature since the superconducting gap becomes smaller. The spectral width at unitarity is comparable, within error bars, to the results of reference \cite{Dyke2024}. 
This result is interesting, since we are comparing a homogeneous and an inhomogeneous system. Therefore our experimental finding seems to suggest that the inhomogeneity alone is not the limiting factor of the line width. To confirm this intuition quantitatively, we have performed a local-density approximation of the Higgs mode in order to account for dephasing effects of many independent oscillators. Such a simple model would lead to significantly broader spectral features than what we observe. Considering these results, it seems that the inhomogeneous density distribution of the trap reduces the observed Higgs frequency but does not lead to a considerable additional broadening of the mode. Therefore, our measurements suggest that a trap-averaged  order parameter dominates the Higgs frequency, as proposed in ref.~\cite{hannibal_persistent_2018} for interaction quenches, however, the measured width is significantly broader than suggested in \cite{hannibal_persistent_2018} and cannot be accounted for by a pure LDA approach as used in ref.~\cite{Dyke2024}.

In conclusion, we have detected the Higgs mode of a strongly interacting Fermi gas using interaction modulation spectroscopy. The observed resonance frequencies follow the interaction dependence of the order parameter as expected if we treat the Higgs frequency in a local-density approximation. However, we observe no temperature dependence of the Higgs mode frequency, which  is in agreement with recent experimental data obtained using a completely different method \cite{Dyke2024}, however, theoretically is not explained and requires future research.

We thank M. Link and K. Gao for contributions in the early phase of this work. This work has been supported by Deutsche Forschungsgemeinschaft through the Cluster of Excellence Matter and Light for Quantum Computing (ML4Q) EXC 2004/1–390534769 and SFB/TR 185 (project B4).

\end{document}